\documentclass[9pt,letterpaper]{article}
\usepackage{spconf,amsmath,amssymb,graphicx,booktabs,cite,float}

\usepackage[T1]{fontenc}

\ninept
\makeatletter
\long\def\@makecaption#1#2{%
  \vskip 4pt
  \setbox\@tempboxa\hbox{#1. #2}%
  \ifdim\wd\@tempboxa>\hsize #1. #2\par\else
    \hbox to\hsize{\hfil\box\@tempboxa\hfil}%
  \fi}
\makeatother

\title{ENVIRONMENT-CONDITIONED DYNAMIC KV CACHING FOR MULTIPATH PARAMETER PREDICTION}
\name{Zhaoyang Li \qquad Yuhao Chen \qquad Qianqian Yang \qquad Zhaoyang Zhang}
\address{College of Information Science and Electronic Engineering, Zhejiang University, Hangzhou, China}

\begin{document}

\maketitle
\begin{abstract}
Multipath parameter prediction provides a structured and physically interpretable representation of wireless propagation for environment-aware communication systems. However, existing Transformer-based methods repeatedly encode the same environment when predicting multiple BS--UE links, resulting in redundant computation. To address this issue, we propose an environment-conditioned dynamic KV caching framework for multipath parameter prediction. Specifically, we decouple environment-dependent computation from BS--UE geometry, enabling environment-side representations and intermediate KV states to be reused across multiple link queries while only the geometry branch is updated. Building on this reusable cache, we further introduce a dynamic compression mechanism that uses environment-side attention scores to retain informative tokens and adaptively allocates a fixed cache budget across Transformer layers. Experiments on the WAIR-D dataset show that the proposed method reduces KV memory from 0.2500 MB to 0.1406 MB, while the delay NMSE increases only slightly from 0.0673 to 0.0678. Code is available at https://github.com/LZhaoyang/kv.
\end{abstract}

\section{Introduction}
Wireless channels in massive MIMO and millimeter-wave systems exhibit pronounced spatial selectivity and sparse yet complex multipath structures, making accurate propagation knowledge essential for reliable communication\cite{rappaport2015wideband,wang2018spatialwideband}. Conventional channel representations, such as channel state information (CSI), covariance matrices, or beam indices, describe the aggregate response of multiple propagation components and therefore obscure the physical relationship between the surrounding environment and individual propagation paths\cite{xu2021covariance,wu2024ckm}. In contrast, a path-wise representation characterizes each dominant multipath component through its gain, delay, angle of departure (AoD), and angle of arrival (AoA). Such a representation is compact and physically interpretable, and can support downstream tasks including channel reconstruction, beam management, localization, and sensing \cite{lin2021joint,gentner2016multipath}. Direct prediction of path-level parameters therefore provides a promising approach to interpretable wireless propagation modeling.

\begin{figure}[!t]
\centering
\includegraphics[width=\columnwidth]{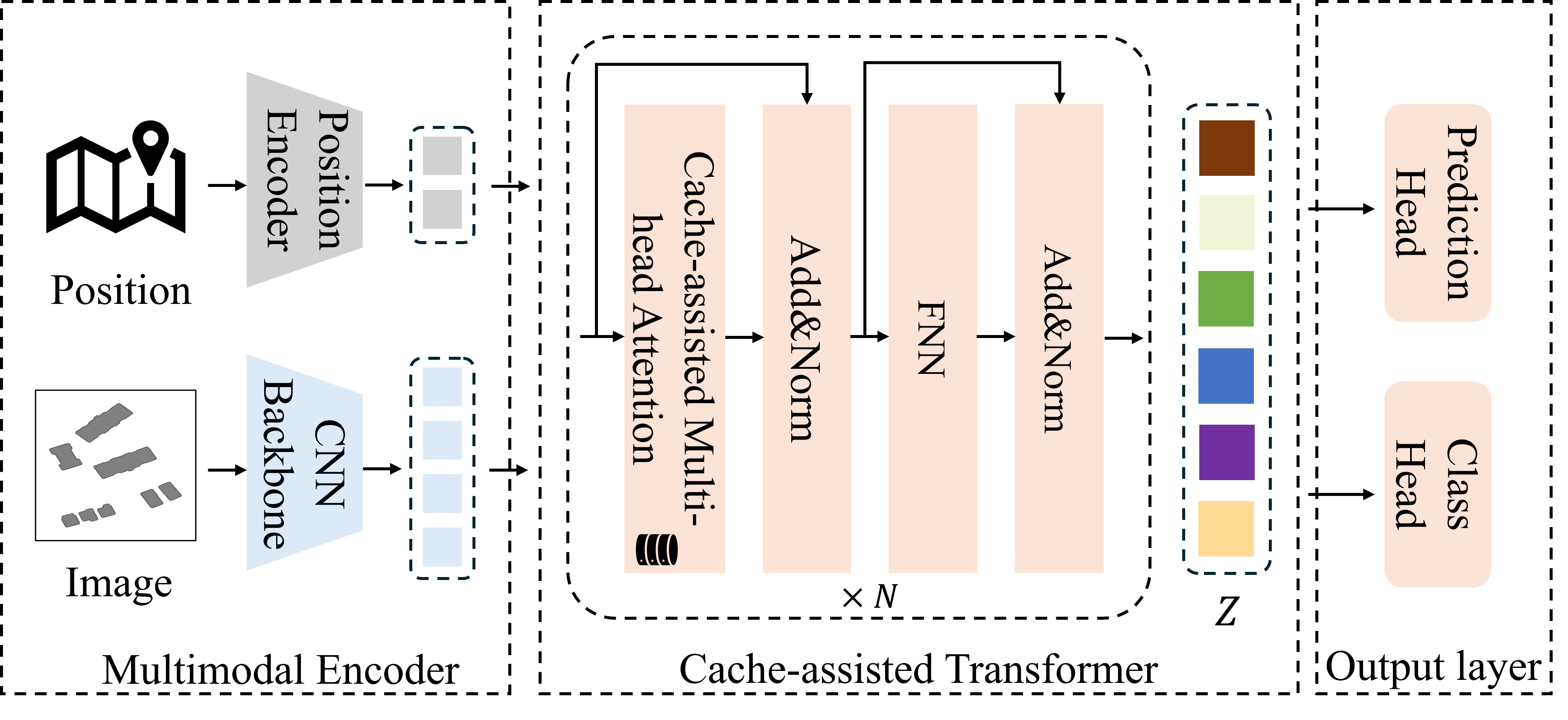}
\caption{Overview of the proposed multipath parameter prediction framework.}
\label{fig:overview}
\end{figure}

Traditional acquisition of multipath parameters mainly relies on deterministic propagation modeling or signal-processing-based parameter estimation. Ray tracing explicitly models interactions between electromagnetic waves and physical objects, but its computational cost increases rapidly with environment complexity \cite{yun2015raytracing,3gpp38901}. Parameter-estimation methods exploit channel sparsity to recover path delays, angles, and gains from channel measurements, but typically require online observations and rely on predefined propagation or sparsity assumptions\cite{haghshenas2024parametric,zhang2024nearfield,
xiao2024movable}. More recently, data-driven approaches have emerged as an attractive alternative. Neural networks have been widely investigated for channel estimation and prediction \cite{he2018beamspace,gao2023wideband,yuan2020aging,
zhou2022highspeed,liu2022spatiotemporal,qin2022partial,
huttunen2023deeptx}, while Transformer-based models have demonstrated strong capability in capturing long-range dependencies and complex spatial-temporal relationships in wireless channels \cite{jiang2022transformer,zhou2024transformer}. Their attention mechanism also provides a natural way to model interactions between environment representations and BS--UE geometry. These properties make Transformer architectures particularly suitable for multipath parameter prediction.

Despite their strong modeling capability, existing Transformer-based predictors can be computationally inefficient when repeatedly queried within the same environment. In practical deployments, the physical environment, including buildings and dominant scatterers, often remains static or changes much more slowly than user locations. As a result, a base station may need to predict multipath parameters for many BS--UE links that share essentially the same environmental context. However, Transformer-based multimodal methods typically process the environmental input and BS--UE geometry jointly for each query, resulting in repeated environment encoding. This redundant computation can be particularly burdensome for base stations and edge devices with limited computational and memory resources.

Inspired by KV caching in Transformer inference\cite{zhang2023h2o,cai2024pyramidkv,zhou2025dynamickv}, we observe that this mechanism naturally fits the above
environment-reuse setting, where previously computed key and value
states can be cached to avoid redundant attention computation.
Motivated by this property, we propose an environment-conditioned dynamic KV
caching mechanism that decouples environment-dependent encoding from
geometry-dependent prediction. Environment representations and
intermediate KV states are computed once and reused across subsequent
link queries. To further reduce cache memory, environment-side attention
scores are used to retain informative environment tokens, while a fixed
cache budget is dynamically allocated across Transformer layers. Experiments on WAIR-D show that the proposed dynamic KV cache reduces KV memory from 0.2500 MB to 0.1406 MB, while the delay NMSE increases only slightly from 0.0673 to 0.0678. To the best of our knowledge, this is the first work to introduce
KV cache reuse into environment-aware multipath parameter prediction.

\section{Method}

\subsection{Multipath Prediction Formulation}
As illustrated in Fig.~\ref{fig:overview}, the proposed framework
takes an environmental image and the corresponding BS--UE geometry as
input, and predicts a set of dominant multipath components.
For environment $e$ and the $m$-th BS--UE link query, let $I_e$
denote the environmental image and $g_{e,m}$ denote the corresponding
BS--UE geometry. The environmental image is encoded into a sequence of
environment tokens, while the BS--UE geometry is mapped to a
geometry query token:
\begin{equation}
\mathbf E_{\mathrm{env}}^{(0)}
=
\Phi_{\mathrm{img}}(I_e),
\qquad
\mathbf q_{\mathrm{geo}}^{(0)}
=
\Phi_{\mathrm{geo}}(g_{e,m}),
\label{eq:input_encoding}
\end{equation}
where
$\mathbf E_{\mathrm{env}}^{(0)}
\in\mathbb{R}^{S_{\mathrm{env}}\times D}$
contains $S_{\mathrm{env}}$ environment tokens with feature dimension
$D$, and
$\mathbf q_{\mathrm{geo}}^{(0)}\in\mathbb{R}^{D}$
represents the BS--UE geometry.

The model predicts a fixed set of at most $K_{\max}$ candidate
multipath components:
\begin{equation}
\hat{\mathcal P}_{e,m}
=
\left\{
\left(
\hat p_k,
\hat{\boldsymbol{\theta}}_k
\right)
\right\}_{k=1}^{K_{\max}},
\label{eq:path_output}
\end{equation}
where $\hat p_k$ denotes the existence confidence of the $k$-th
candidate path and
$\hat{\boldsymbol{\theta}}_k\in\mathbb{R}^{6}$
contains its propagation delay, two-dimensional AoA,
two-dimensional AoD, and path gain. In our implementation,
$K_{\max}=16$. 

\begin{figure}[!t]
\centering
\includegraphics[width=0.85\columnwidth]{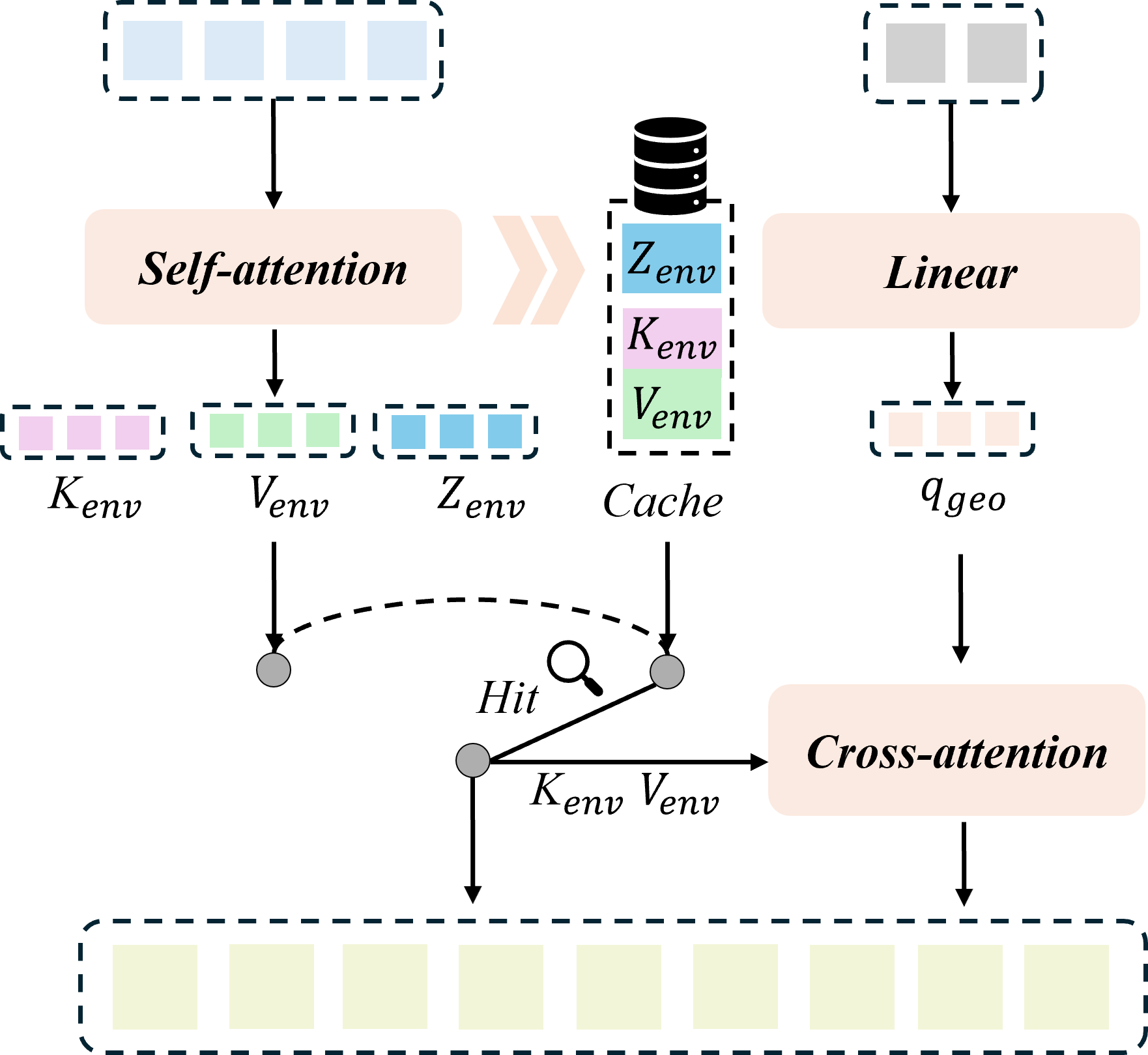}
\caption{KV Caching for Environment Reuse.}
\label{fig:reuse}
\end{figure}

\subsection{KV Caching for Environment Reuse}

For a fixed environment, multiple BS--UE queries share the same
environment representation while only the BS--UE geometry changes.
To exploit this environment-level reusability, we introduce a
decoupled inference scheme with KV cache reuse, as illustrated in
Fig.~\ref{fig:reuse}. 

For an $L$-layer Transformer, where $L$ denotes the number of encoder
layers, the environment-side query, key, and value states at layer
$\ell$ are
\begin{equation}
\left(
\mathbf Q_{\mathrm{env}}^{(\ell)},
\mathbf K_{\mathrm{env}}^{(\ell)},
\mathbf V_{\mathrm{env}}^{(\ell)}
\right)
=
\operatorname{Proj}_{QKV}^{(\ell)}
\left(
\mathbf E_{\mathrm{env}}^{(\ell-1)}
\right),
\label{eq:env_qkv}
\end{equation}
where $\operatorname{Proj}_{QKV}^{(\ell)}(\cdot)$ denotes the
query--key--value projections at layer $\ell$. The environment tokens
perform self-attention using these states and are then updated by the
standard residual and feed-forward operations. 

The reusable environment cache is defined as
\begin{equation}
\mathcal C_{\mathrm{env}}
=
\left(
\mathbf E_{\mathrm{env}}^{(L)},
\left\{
\left(
\mathbf K_{\mathrm{env}}^{(\ell)},
\mathbf V_{\mathrm{env}}^{(\ell)}
\right)
\right\}_{\ell=1}^{L}
\right),
\label{eq:env_cache}
\end{equation}
where $\mathcal C_{\mathrm{env}}$ contains the final environment
representation and all layer-wise environment-side KV states.

For the first query in a new environment, the image encoder and
environment branch are evaluated to construct
$\mathcal C_{\mathrm{env}}$. For subsequent queries, these computations
are skipped and only the geometry-dependent branch is recomputed.

At layer $\ell$, the geometry query token is projected into the
geometry-side query, key, and value states as
$(\mathbf q_{\mathrm{geo}}^{(\ell)},
\mathbf k_{\mathrm{geo}}^{(\ell)},
\mathbf v_{\mathrm{geo}}^{(\ell)})
=
\operatorname{Proj}_{QKV}^{(\ell)}
(\mathbf q_{\mathrm{geo}}^{(\ell-1)})$.
Its key and value states are concatenated with the cached environment
states:
\begin{equation}
\mathbf K_{\mathrm{all}}^{(\ell)}
=
\left[
\mathbf K_{\mathrm{env}}^{(\ell)};
\mathbf k_{\mathrm{geo}}^{(\ell)}
\right],
\quad
\mathbf V_{\mathrm{all}}^{(\ell)}
=
\left[
\mathbf V_{\mathrm{env}}^{(\ell)};
\mathbf v_{\mathrm{geo}}^{(\ell)}
\right].
\label{eq:kv_concat}
\end{equation}
The geometry-dependent attention is then computed as
\begin{equation}
\mathbf z_{\mathrm{geo}}^{(\ell)}
=
\operatorname{Attn}
\left(
\mathbf q_{\mathrm{geo}}^{(\ell)},
\mathbf K_{\mathrm{all}}^{(\ell)},
\mathbf V_{\mathrm{all}}^{(\ell)}
\right),
\label{eq:cache_attn}
\end{equation}
where $\operatorname{Attn}(\cdot)$ denotes the standard scaled
dot-product attention. After the residual and feed-forward operations,
the updated geometry token is propagated to the next layer.

\subsection{Environment-Conditioned Dynamic KV Compression}

\begin{figure}[!t]
\centering
\includegraphics[width=\columnwidth]{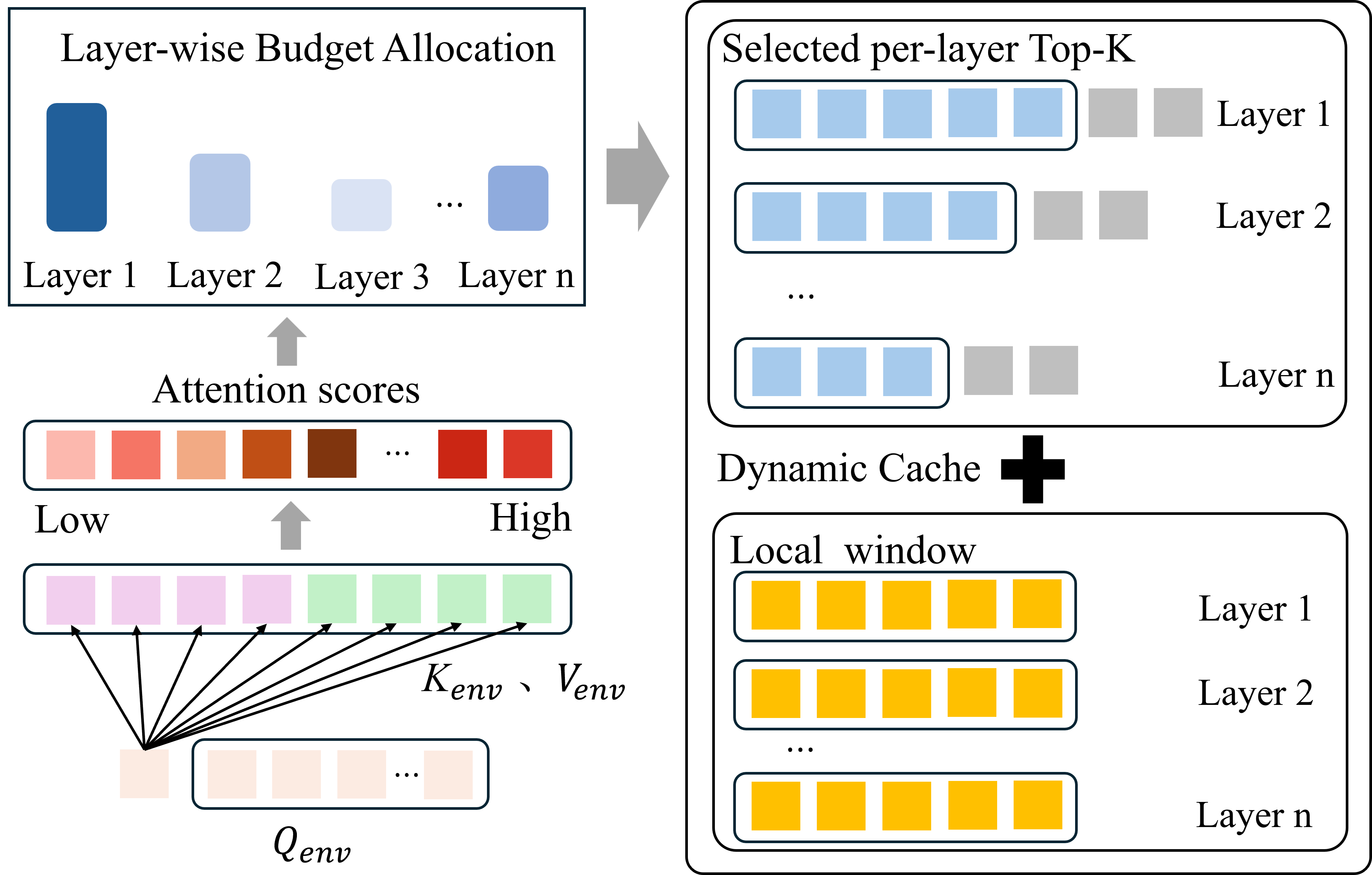}
\caption{Environment-Conditioned Dynamic KV Compression.}
\label{fig:dynamic}
\end{figure}

Although environment reuse avoids repeated environment encoding, the
full cache still stores KV states for all $S_{\mathrm{env}}$ tokens at
every Transformer layer. We therefore combine within-layer token
selection with cross-layer budget allocation, as shown in
Fig.~\ref{fig:dynamic}. Since token selection depends only on
environment-side attention, the compressed cache remains reusable
across geometry queries.

At layer $\ell$,
$\mathbf Q_{\mathrm{env}}^{(\ell)},
\mathbf K_{\mathrm{env}}^{(\ell)},$
and $\mathbf V_{\mathrm{env}}^{(\ell)}
\in\mathbb{R}^{H\times S_{\mathrm{env}}\times d_h}$,
where $H$ denotes the number of attention heads and $d_h$ the
dimension of each head. The last $W$ environment tokens form a
retained local window, where $W$ is the window size, while the
preceding $S_{\mathrm{env}}-W$ tokens constitute the compressible
historical region.

For head $h$, the attention weights from the retained window to all
environment tokens are
\begin{equation}
\mathbf A_h^{(\ell)}
=
\operatorname{Softmax}
\left(
\frac{
\mathbf Q_{\mathrm{env},h,-W:}^{(\ell)}
\left(\mathbf K_{\mathrm{env},h}^{(\ell)}\right)^T
}{
\sqrt{d_h}
}
\right),
\label{eq:window_attn}
\end{equation}
where $\mathbf A_h^{(\ell)}$ denotes the corresponding attention
matrix. The importance of historical token $t$ is accumulated over the
window queries and smoothed by one-dimensional pooling:
\begin{equation}
r_{h,t}^{(\ell)}
=
\sum_{w=1}^{W}A_{h,w,t}^{(\ell)},
\qquad
\bar{\mathbf r}_{h}^{(\ell)}
=
\operatorname{Pool}_{\kappa}
\left(\mathbf r_h^{(\ell)}\right),
\label{eq:token_score}
\end{equation}
where $t=1,\ldots,S_{\mathrm{env}}-W$, $\mathbf r_h^{(\ell)}$
collects the historical-token scores, and $\kappa$ denotes the pooling
kernel size.

Let $C_{\max}$ denote the target average cache capacity per layer.
The historical-token budget $B_0$ and temporary candidate budget
$B_{\mathrm{tmp}}$ are
\begin{equation}
B_0=C_{\max}-W,
\qquad
B_{\mathrm{tmp}}
=
\min
\left(
\left\lfloor\rho_{\max}B_0\right\rfloor,
S_{\mathrm{env}}-W
\right),
\label{eq:tmp_budget}
\end{equation}
where $\rho_{\max}$ controls the candidate-pool size. For each head,
the temporary candidate-index set is
\begin{equation}
\mathcal I_{h,\mathrm{tmp}}^{(\ell)}
=
\operatorname{TopKIdx}
\left(
\bar{\mathbf r}_{h}^{(\ell)},
B_{\mathrm{tmp}}
\right),
\label{eq:tmp_indices}
\end{equation}
where $\operatorname{TopKIdx}(\cdot)$ returns the indices of the
largest-scoring entries.

The corresponding temporary environment-side KV states are
\begin{align}
\mathbf K_{\mathrm{env},h,\mathrm{tmp}}^{(\ell)}
&=
\left[
\mathbf K_{\mathrm{env},h,
\mathcal I_{h,\mathrm{tmp}}^{(\ell)}}^{(\ell)};
\mathbf K_{\mathrm{env},h,-W:}^{(\ell)}
\right],
\\
\mathbf V_{\mathrm{env},h,\mathrm{tmp}}^{(\ell)}
&=
\left[
\mathbf V_{\mathrm{env},h,
\mathcal I_{h,\mathrm{tmp}}^{(\ell)}}^{(\ell)};
\mathbf V_{\mathrm{env},h,-W:}^{(\ell)}
\right].
\label{eq:tmp_cache}
\end{align}

To further adapt the cache size across layers, let $\mathcal G$ denote
the set of processed Transformer layers. Their pooled attention scores
are jointly ranked under the total historical-token budget
$B_0H|\mathcal G|$:
\begin{equation}
\mathcal I_{\mathrm{global}}
=
\operatorname{TopKIdx}
\left(
\left[
\left\{
\bar{\mathbf r}_{h}^{(\ell)}
\right\}_{h=1}^{H}
\right]_{\ell\in\mathcal G},
B_0H|\mathcal G|
\right),
\label{eq:global_selection}
\end{equation}
where $\mathcal I_{\mathrm{global}}$ denotes the globally retained
score entries. Let $c_\ell$ denote the number of these entries belonging
to layer $\ell$. The historical-token budget assigned to that layer is
\begin{equation}
B_\ell
=
\left\lfloor
B_0|\mathcal G|
\frac{c_\ell}{\sum_{j\in\mathcal G}c_j}
\right\rceil,
\qquad
\sum_{\ell\in\mathcal G}B_\ell
=
B_0|\mathcal G|.
\label{eq:layer_budget}
\end{equation}
A residual correction after rounding preserves the total budget.

Let $\mathcal I_h^{(\ell)}$ denote the top-$B_\ell$ historical indices
selected from $\mathcal I_{h,\mathrm{tmp}}^{(\ell)}$. The final
compressed environment-side states are
\begin{align}
\widetilde{\mathbf K}_{\mathrm{env},h}^{(\ell)}
&=
\left[
\mathbf K_{\mathrm{env},h,\mathcal I_h^{(\ell)}}^{(\ell)};
\mathbf K_{\mathrm{env},h,-W:}^{(\ell)}
\right],\\
\widetilde{\mathbf V}_{\mathrm{env},h}^{(\ell)}
&=
\left[
\mathbf V_{\mathrm{env},h,\mathcal I_h^{(\ell)}}^{(\ell)};
\mathbf V_{\mathrm{env},h,-W:}^{(\ell)}
\right].
\label{eq:compressed_kv}
\end{align}
For subsequent BS--UE queries, these compressed states replace the full
environment-side KV states in \eqref{eq:cache_attn}.

The corresponding KV-memory costs scale as
\begin{equation}
\mathcal M_{\mathrm{full}}
\propto
2LHS_{\mathrm{env}}d_h,
\qquad
\mathcal M_{\mathrm{dyn}}
\propto
2H
\sum_{\ell=1}^{L}(B_\ell+W)d_h,
\label{eq:cache_memory}
\end{equation}
where $\mathcal M_{\mathrm{full}}$ and $\mathcal M_{\mathrm{dyn}}$
denote the KV-memory costs of the full and dynamic caches, respectively.

\section{Experiments}

\begin{figure*}[!t]
\centering
\begin{minipage}[b]{0.245\textwidth}
\centering
\includegraphics[width=\linewidth]{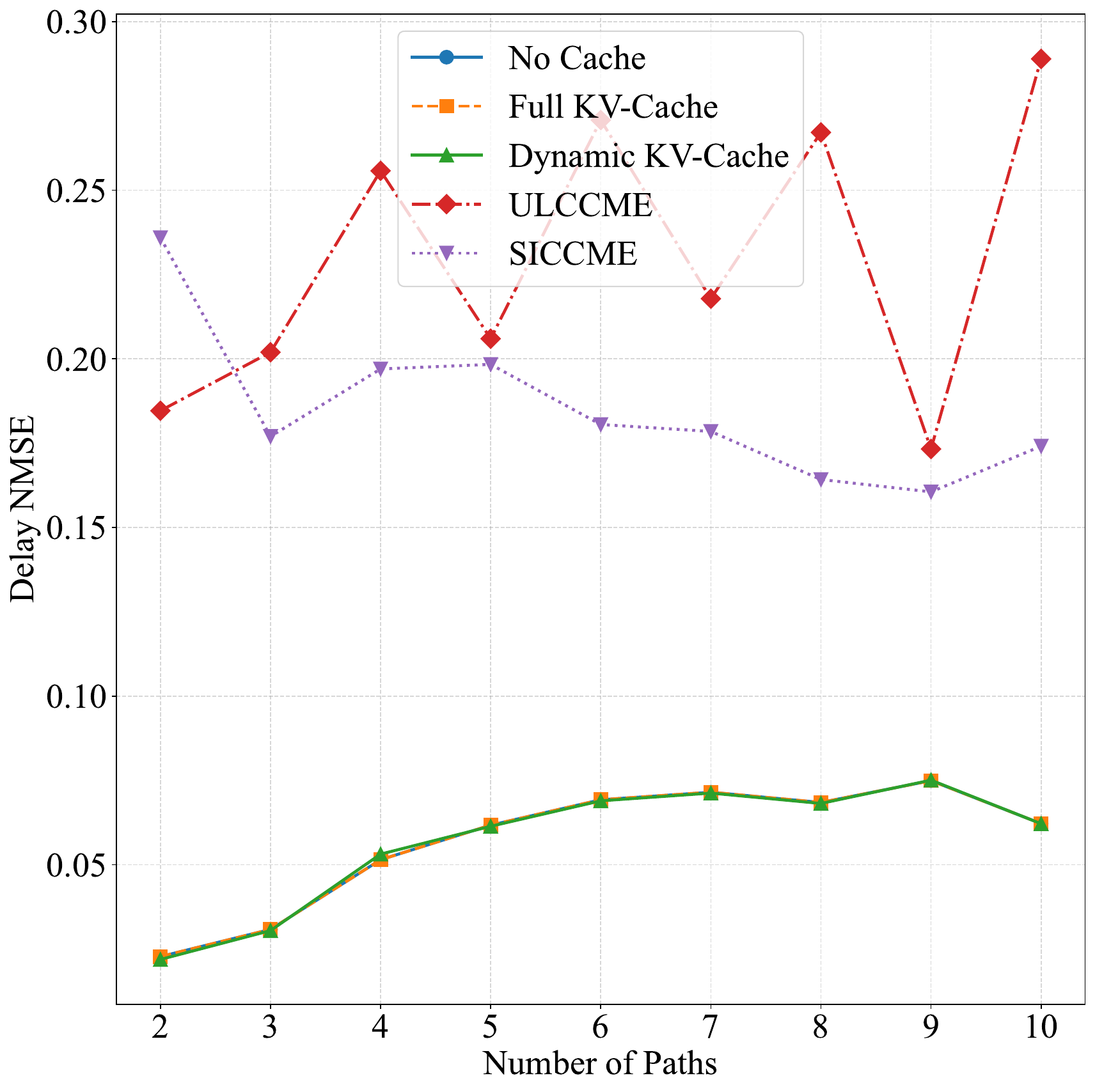}\\[-1mm]
(a) Delay NMSE
\end{minipage}\hspace{-1mm}
\begin{minipage}[b]{0.245\textwidth}
\centering
\includegraphics[width=\linewidth]{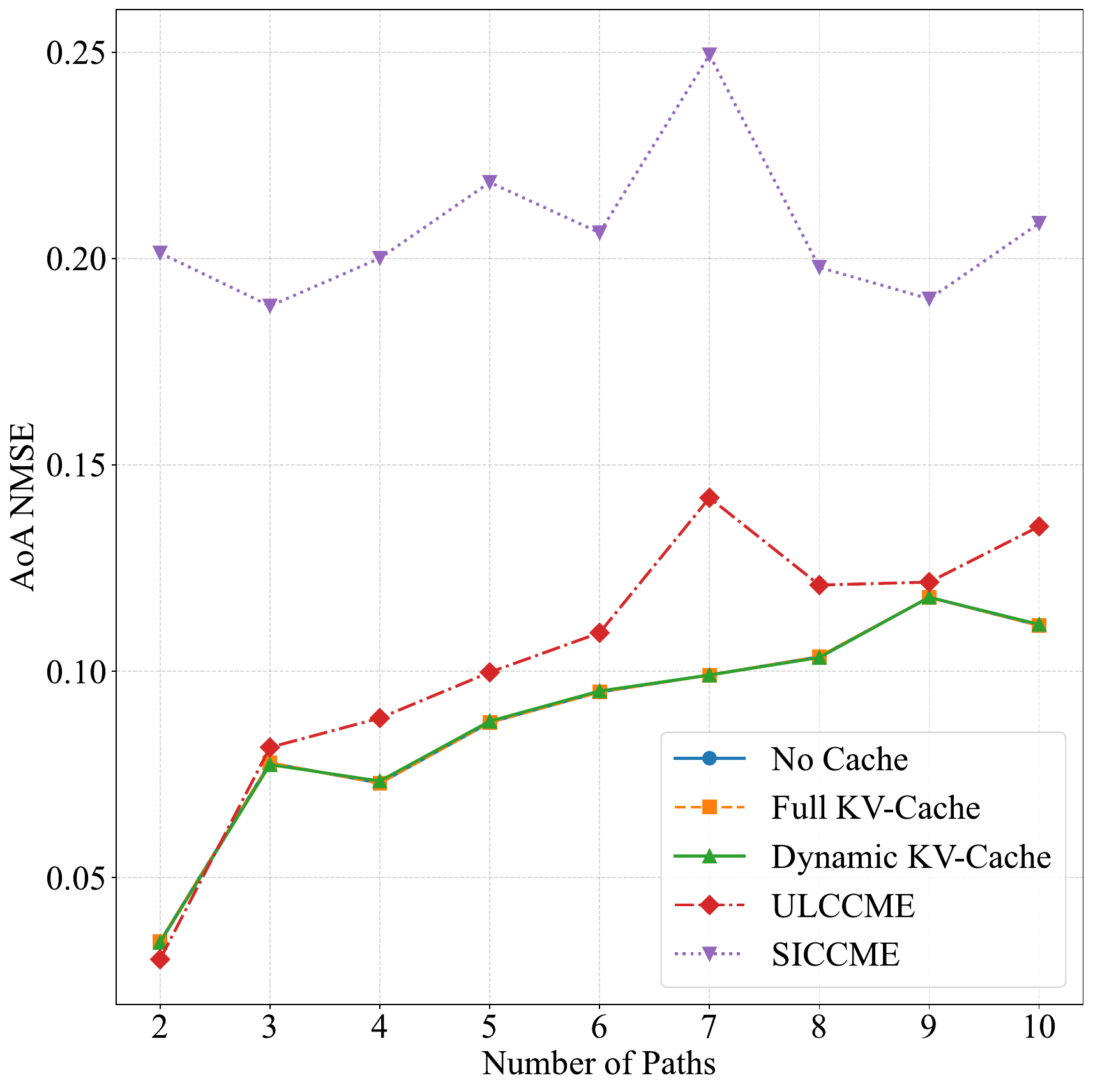}\\[-1mm]
(b) AoA NMSE
\end{minipage}\hspace{-1mm}
\begin{minipage}[b]{0.245\textwidth}
\centering
\includegraphics[width=\linewidth]{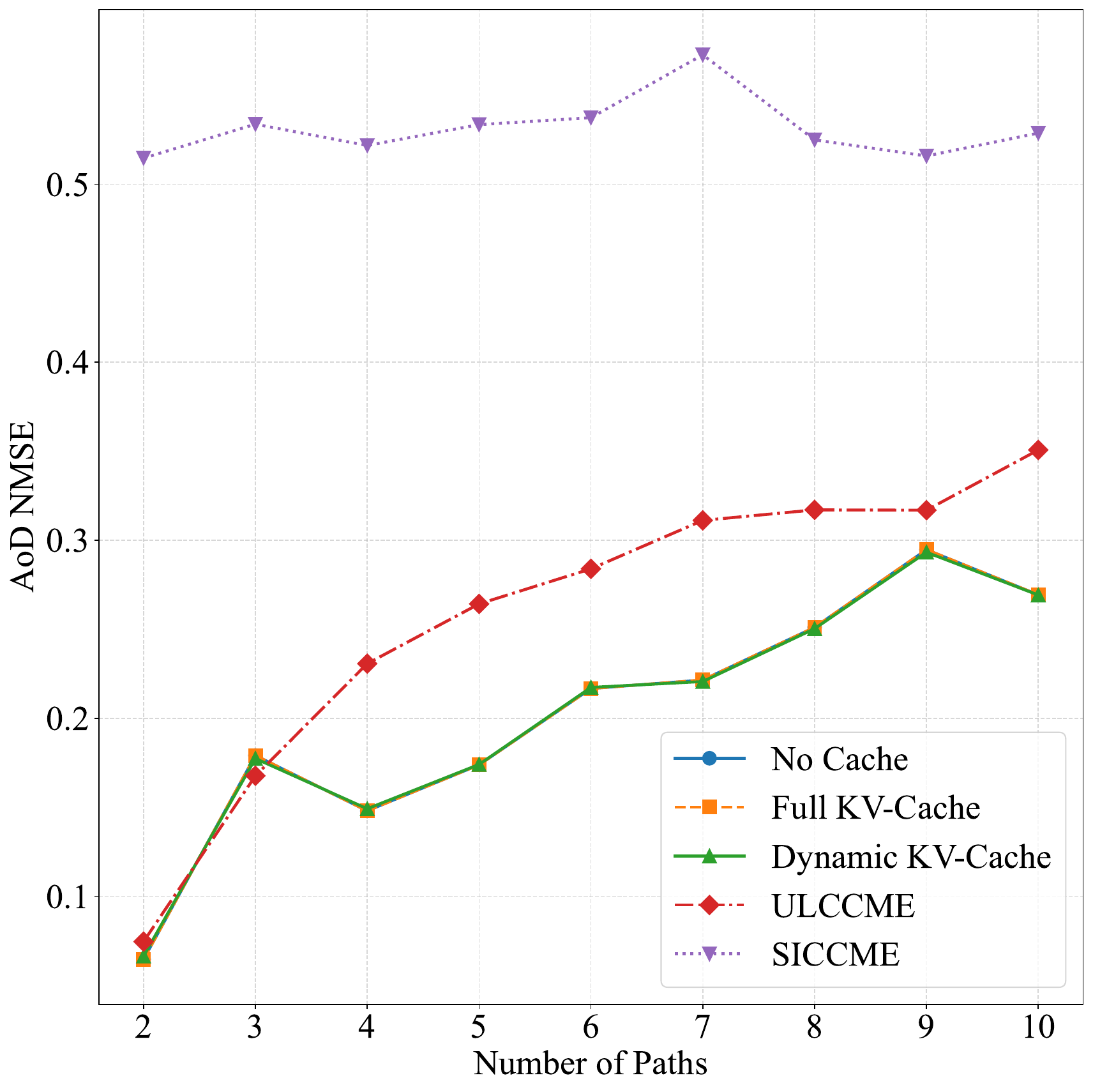}\\[-1mm]
(c) AoD NMSE
\end{minipage}\hspace{-1mm}
\begin{minipage}[b]{0.245\textwidth}
\centering
\includegraphics[width=\linewidth]{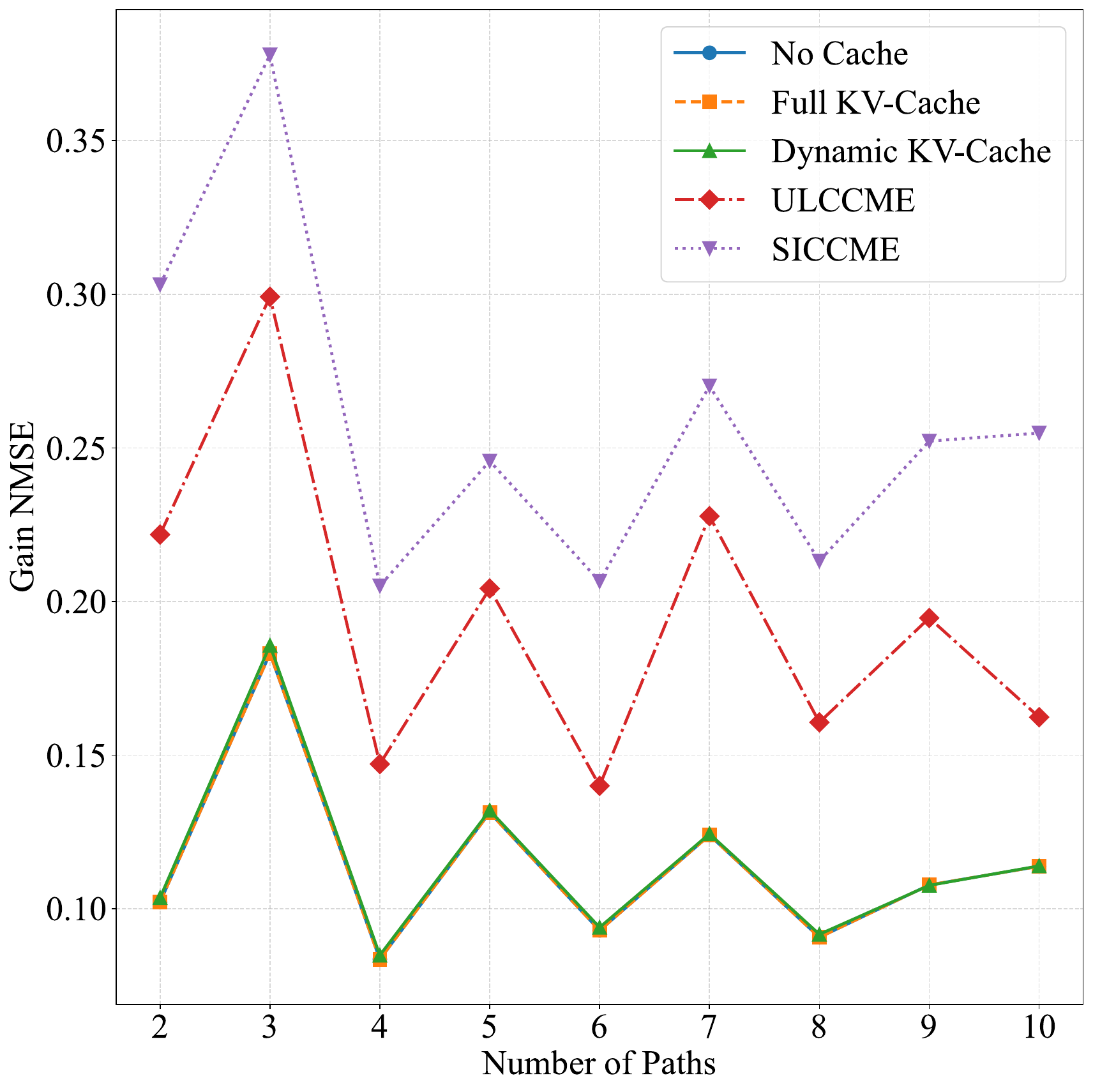}\\[-1mm]
(d) Gain NMSE
\end{minipage}
\caption{NMSE comparison under different prediction and cache settings.}
\label{fig:cachecurves}
\end{figure*}

\subsection{Experimental Setup}

Experiments use WAIR-D \cite{huangfu2022waird}. In Scene~1, each
environment contains five BSs and 30 UEs; the carrier frequency is
28~GHz, and 64 of 384 subcarriers are sampled. The proposed predictor
uses four Transformer encoder layers, eight attention heads, a hidden
dimension of 512, and 16 output slots. For comparison, ULCCME and SICCME from \cite{xu2021covariance} are
adapted to the same multipath parameter prediction task using
geometry-only and image-plus-geometry inputs, respectively. Both are
non-Transformer baselines, and their output layers are adapted to the
same path targets. All experiments are conducted on a single NVIDIA RTX A6000 GPU.

\subsection{Transformer-Based Prediction Performance}

\begin{figure}[!t]
\centering
\includegraphics[width=0.85\columnwidth]{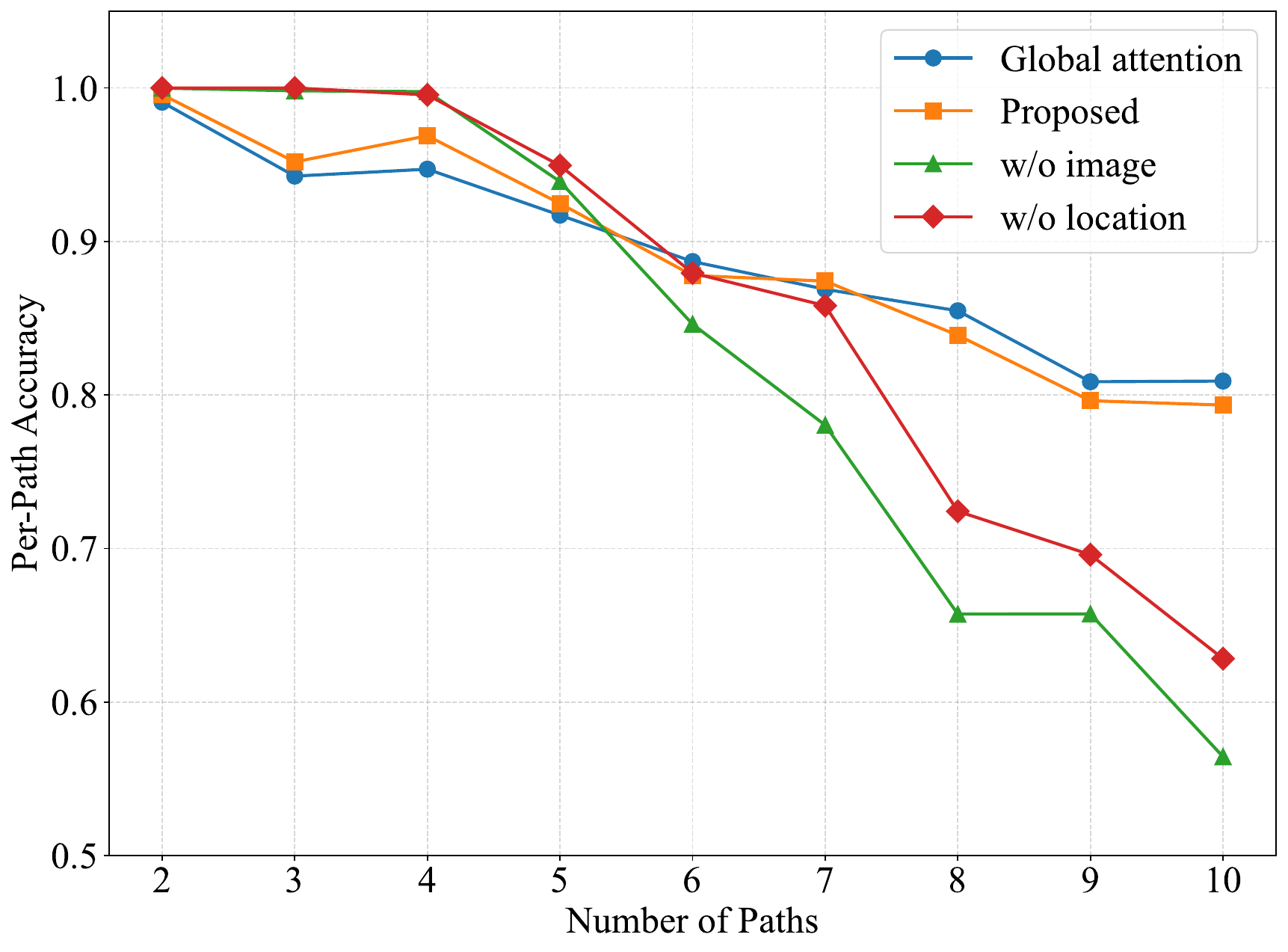}
\caption{Per-path accuracy on Scene~1.}
\label{fig:ablation}
\end{figure}

As shown in Fig.~\ref{fig:cachecurves}, the proposed Transformer
predictor without cache reuse achieves lower delay, AoA, AoD, and gain
NMSE than the adapted ULCCME and SICCME baselines across most
path-count settings. Since the compared models predict the same
path-level targets, these results demonstrate the advantage of
Transformer-based interaction modeling for capturing the complex
relationship between environmental information and BS--UE geometry. The ablation results in Fig.~\ref{fig:ablation} further show that both
input modalities are important for complex multipath prediction. At ten
paths, the proposed model achieves a per-path accuracy of 0.793,
compared with 0.564 without the environmental image and 0.628 without
geometry. Moreover, the proposed decoupled Transformer remains close to
the global-attention variant in per-path accuracy, indicating that the
cache-compatible decoupling preserves strong prediction capability while
enabling subsequent KV reuse.

\subsection{KV Cache Reuse and Compression Gains}

Having established the prediction capability of the Transformer
architecture, we next isolate the effect of KV caching by comparing the
same decoupled predictor under no-cache, full KV cache, and dynamic KV
cache inference. As shown in Fig.~\ref{fig:cachecurves}, the three modes produce nearly
overlapping NMSE curves for all four multipath parameters. Full KV
cache reuse exactly reproduces the uncached NMSE values in
Table~\ref{tab:cache}, confirming that reusing environment-side KV
states does not alter the prediction itself. Meanwhile, the mean online
latency decreases from 9.20~ms to 7.16~ms, demonstrating the direct
computational benefit of avoiding repeated environment encoding. Dynamic KV compression further reduces the KV memory from
0.2500~MB to 0.1406~MB, corresponding to a 43.75\% reduction, while
introducing only marginal changes in the four NMSE metrics. Therefore,
KV caching provides two complementary deployment benefits: full KV
reuse removes redundant environment-side computation, while dynamic KV
compression substantially reduces the memory required to store reusable
states. 

\begin{table}[!t]
\centering
\caption{Performance under different KV cache
modes on Scene~1.}
\label{tab:cache}

\setlength{\tabcolsep}{1.5pt}
\renewcommand{\arraystretch}{0.95}

\begin{tabular}{@{}lrrr@{}}
\toprule
Metric & No Cache & Full KV & Dynamic KV \\
\midrule
Delay NMSE       & 0.0673 & 0.0673 & 0.0678 \\
AoA NMSE         & 0.0875 & 0.0875 & 0.0877 \\
AoD NMSE         & 0.2097 & 0.2097 & 0.2105 \\
Gain NMSE        & 0.1108 & 0.1108 & 0.1116 \\
Mean online (ms) & 9.20   & 7.16   & 10.31  \\
P90 online (ms)  & 9.73   & 7.23   & 10.35  \\
KV memory (MB)   & 0      & 0.2500 & 0.1406 \\
\bottomrule
\end{tabular}

\end{table}

\subsection{Dynamic KV Cache Capacity}

Table~\ref{tab:capacity} further evaluates the effect of the cache
capacity with $W=4$. Reducing $C_{\max}$ from 12 to 6 decreases the
KV memory from 0.1875~MB to 0.0938~MB, while the reported NMSE
changes only from 0.0677 to 0.0683. At $C_{\max}=6$, the dynamic KV
cache reduces memory by 62.50\% relative to the full cache. These
results show that the cache capacity provides a controllable
memory--accuracy tradeoff for deployment under different memory
constraints.

\begin{table}[!t]
\centering
\caption{Dynamic KV cache capacity comparison with $W=4$.}
\label{tab:capacity}

\setlength{\tabcolsep}{1.5pt}
\renewcommand{\arraystretch}{0.95}

\begin{tabular}{@{}rrrrr@{}}
\toprule
$C_{\max}$ & NMSE & Online & Memory & Reduction \\
           &      & (ms)   & (MB)   &           \\
\midrule
6  & 0.0683 & 7.57  & 0.0938 & 62.50\% \\
9  & 0.0678 & 10.31 & 0.1406 & 43.75\% \\
12 & 0.0677 & 12.18 & 0.1875 & 25.00\% \\
\bottomrule
\end{tabular}

\end{table}

\subsection{Cross-Scenario Generalization}

\begin{figure}[!t]
\centering
\includegraphics[width=0.85\columnwidth]{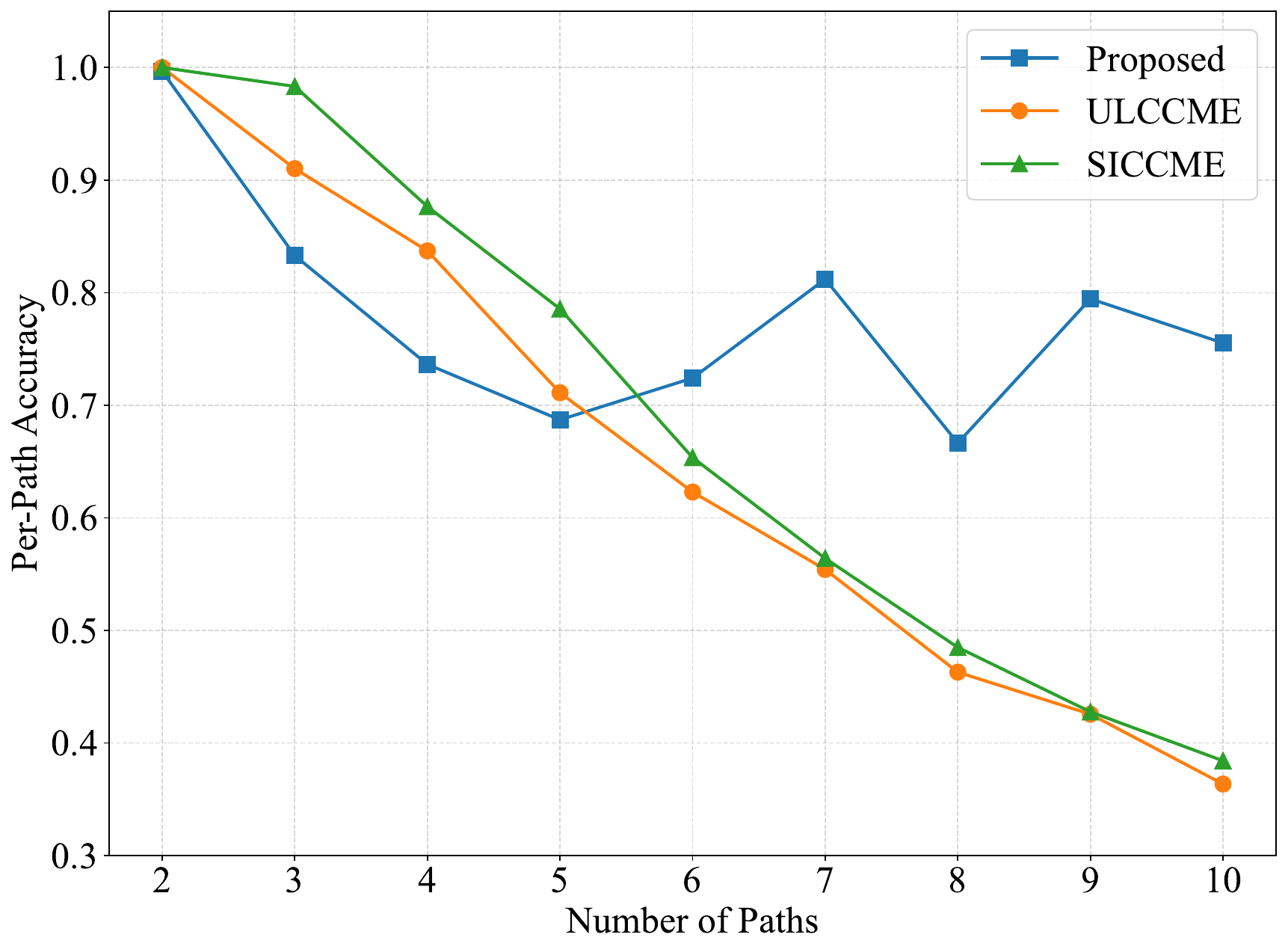}
\caption{Cross-scenario per-path accuracy on Scenario~2.}
\label{fig:transfer}
\end{figure}

Finally, we evaluate whether the learned Transformer representation
generalizes beyond the training scenario. The model trained on
Scenario~1 is directly evaluated on Scenario~2 without fine-tuning.
Scenario~2 contains 100 environments, each with one BS and 10,000
densely distributed UEs, introducing both unseen environments and a
substantially different UE-location distribution.
As shown in Fig.~\ref{fig:transfer}, the proposed model maintains
substantially higher per-path accuracy than the adapted non-Transformer
baselines under more complex multipath conditions. For eight, nine, and
ten paths, the proposed model achieves accuracies of 0.666, 0.795, and
0.755, respectively. These results further demonstrate the superior generalization of the Transformer-based method.

\begin{table}[!t]
\centering
\caption{Per-path accuracy on Scenario~2.}
\label{tab:transfer}
\setlength{\tabcolsep}{3pt}
\begin{tabular}{rrrr}
\toprule
Paths & Proposed & ULCCME & SICCME\\
\midrule
8  & 0.666 & 0.463 & 0.485\\
9  & 0.795 & 0.426 & 0.428\\
10 & 0.755 & 0.364 & 0.384\\
\bottomrule
\end{tabular}
\end{table}

\section{Conclusion}
In this paper, we proposed an environment-conditioned dynamic KV caching
framework for repeated multipath parameter prediction in
environment-reuse scenarios. The proposed method decoupled
environment-dependent computation from geometry-dependent prediction,
allowing environment-side representations and layer-wise KV states to
be computed once and reused across multiple BS--UE queries. To further
reduce cache overhead, a dynamic KV cache compression mechanism was
developed to select informative environment tokens from environment-side
attention and adaptively allocate the cache budget across Transformer
layers. Experimental results on WAIR-D showed that the proposed dynamic KV
cache reduced KV memory by 43.75\%, while increasing the delay NMSE
by only 0.0005. These results demonstrated an effective
memory--accuracy tradeoff for repeated multipath parameter prediction.

\small
\bibliographystyle{IEEEbib}
\bibliography{references}
\end{document}